\documentclass[11pt]{cernrep}
\usepackage{graphicx}
\usepackage{axodraw}
\usepackage{here}

\begin{document}
\bibliographystyle{unsrt}
\renewcommand{\thefootnote}{\fnsymbol{footnote}}

\begin{flushright}
hep-ph/0401060\\
LU TP 04--04\\
January 2004
\end{flushright}

\vspace{10mm}

\noindent{\Large\bf MULTIPLE INTERACTIONS AND BEAM REMNANTS%
\footnote{submitted to the proceedings of the Workshop on 
Physics at TeV Colliders, Les Houches, France, 
26 May -- 6 June 2003}}\\[8mm]
\textit{T. Sj\"ostrand and P. Skands}\\
Department of Theoretical Physics, Lund University, 
S\"olvegatan 14A, S-223 62 Lund, Sweden

\section{INTRODUCTION} 

Hadrons are composite systems of quarks and gluons. A direct 
consequence is the possibility to have hadron--hadron collisions 
in which several distinct pairs of partons collide with each other: 
multiple interactions, a.k.a.\ multiple scatterings. 
At first glance, the divergence of the perturbative $t$-channel 
one-gluon-exchange graphs in the $p_{\perp} \to 0$ limit implies an
infinity of interactions per event. However, the perturbative framework 
does not take into account screening from the fact that a hadron is in 
an overall colour singlet state. Therefore an effective cutoff 
$p_{\perp\mathrm{min}}$ of the order of one to a few GeV is introduced, 
representing an inverse colour correlation distance inside the hadron.
For realistic $p_{\perp\mathrm{min}}$ values most inelastic events in 
high-energy hadronic collisions should then contain several 
perturbatively calculable interactions, in addition to whatever 
nonperturbative phenomena may be present.   

Although most of this activity is not hard enough to play a significant 
role in the description of high--$p_{\perp}$ jet physics, it can be
responsible for a large fraction of the total multiplicity (and large
\emph{fluctuations} in it), for semi-hard (mini-)jets in the event, for 
the details of jet profiles and for the jet pedestal effect, leading to 
random as well as systematic shifts in the jet energy scale. Thus, a
good understanding of multiple interactions would seem prerequisite 
to carrying out precision studies involving jets and/or the underlying 
event in hadronic collisions. 

In an earlier study \cite{Sjostrand:1987su}, it was argued that \emph{all} 
the underlying event activity is triggered by the multiple interactions 
mechanism. However, while the origin of underlying events is thus 
assumed to be perturbative, many nonperturbative aspects still need to be
considered and understood:\\
\textit{(i)} What is the detailed mechanism and functional form of the 
dampening of the perturbative cross section at small $p_{\perp}$?
(Certainly a smooth dampening is more realistic than a sharp 
$p_{\perp\mathrm{min}}$ cutoff.)\\ 
\textit{(ii)} Which energy dependence would this mechanism have?\\
\textit{(iii)} How is the internal structure of the proton reflected in an
impact-parameter-dependent multiple interactions rate, as manifested
e.g.\ in jet pedestal effects?\\
\textit{(iv)} How can the set of colliding partons from a hadron be
described in terms of correlated multiparton distribution functions
of flavours and longitudinal momenta?\\
\textit{(v)} How does a set of initial partons at some low perturbative 
cutoff scale, `initiators', evolve into such a set of colliding partons?
(Two colliding partons could well have a common initiator.)
Is standard DGLAP evolution sufficient, or must BFKL/CCFM effects be 
taken into account?\\ 
\textit{(vi)} How would the set of initiators correlate with the flavour 
content of, and the longitudinal momentum sharing inside, the left-behind 
beam remnant?\\
\textit{(vii)} How are the initiator and remnant partons correlated by 
confinement effects (`primordial $k_{\perp}$')?\\
\textit{(viii)} How are all produced partons, both the 
interacting and the beam-remnant ones, 
correlated in colour? Is the large number-of-colours limit 
relevant, wherein partons can be hooked up into strings (with quarks as 
endpoints and gluons as intermediate kinks) representing a linear 
confinement force \cite{Andersson:1983ia}?\\
\textit{(ix)} How is the original baryon number of an incoming proton 
reflected in the colour topology?\\ 
\textit{(x)} To what extent would a framework with independently 
fragmenting  string systems, as defined from the colour topology, be 
modified by the space--time overlap of several strings?

Needless to say, we should not expect to find a perfect solution to any
of these issues, but only successively improved approximations. The 
framework in \cite{Sjostrand:1987su} is very primitive in a number of 
respects. Nevertheless, it has turned out to be quite successful. Thus 
the \textsc{Pythia} Tune A of R.D.~Field \cite{TuneAField} is capable 
of describing a host of jet and minimum-bias event data at the Tevatron. 
The model appears inadequate to fully describe correlations and 
fluctuations, however, and we would expect a poor performance for several 
topics not yet studied experimentally.

In particular, only very simple beam remnant structures could 
technically be dealt with in \cite{Sjostrand:1987su}. One recent 
development was the extension of the standard Lund string framework
\cite{Andersson:1983ia} to include a junction fragmentation description 
\cite{Sjostrand:2002ip} that allows the hadronization of nontrivial 
colour topologies containing non-zero baryon number. In the context of 
multiple interactions, this improvement means that almost arbitrarily 
complicated baryon beam remnants may now be dealt with, hence many of 
the restrictions present in the old model are no longer necessary.
 
Here, we report on the development of a new model for the flavour-, 
colour-, and momentum-correlated partonic structure involved in a 
hadron--hadron collision, i.e.\ partly addressing several of the points 
above. We first present the main work on flavour and momentum space 
correlations, and thereafter separately the very thorny issue of colour 
correlations, before concluding. A more complete description of the model, 
also including references to experimental data and other theoretical 
ideas, and with comments on the all the issues, may be found in 
\cite{multintinprep}. A toy model study of the first two points 
is found in \cite{Dischler:2000pk}. The \textsc{Pythia} manual
\cite{Sjostrand:2003wg} contains some complementary information.
                   
\section{CORRELATED PARTON DENSITIES}

Consider a hadron undergoing multiple interactions in a collision. 
Such an object should be described by multi-parton densities, 
giving the joint probability of simultaneously finding $n$ partons with 
flavours $f_1,\ldots,f_n$, carrying momentum fractions $x_1,\ldots,x_n$ 
inside the hadron, when probed by interactions at scales 
$Q_1^2,\ldots,Q_n^2$. However, we are nowhere near having sufficient 
experimental information to pin down such distributions. Therefore, and
wishing to make maximal use of the information that we \emph{do} have, namely
the standard one-parton-inclusive parton densities, we propose the following
strategy. 

As described in \cite{Sjostrand:1987su}, the interactions may be generated in
an ordered sequence of falling $p_{\perp}$. For the hardest interaction, all
smaller $p_{\perp}$ scales may be effectively integrated out of the (unknown) 
fully correlated  distributions, leaving an object described by the standard
one-parton distributions, by definition. For the second and subsequent
interactions, again all lower--$p_{\perp}$ scales can be integrated out, but 
the correlations with the first cannot, and so on. 
Thus, we introduce modified parton densities, that correlate the $i$'th
interaction and its shower evolution to what happened in the $i-1$ previous
ones. 

The first and most trivial observation is that each interaction $i$
removes a momentum fraction $x_i$ from the hadron remnant. Already in
\cite{Sjostrand:1987su} this momentum loss was taken into account by assuming
a simple scaling ansatz for the parton distributions, $f(x) \to f(x/X)/X$,
where $X = 1 - \sum_{i=1}^n x_i$ is the momentum remaining in the beam hadron
after the $n$ first interactions. Effectively, the PDF's are simply `squeezed' 
into the range $x\in[0,X]$. 

Next, for a given baryon, the valence distribution of flavour $f$ after $n$ 
interactions, $q_{f\mathrm{v} n}(x,Q^2)$, should integrate to the number
$N_{f\mathrm{v} n}$ of valence quarks of flavour $f$ remaining in the hadron 
remnant. This rule may be enforced by scaling the original distribution down, 
by the ratio of remaining to original valence quarks 
$N_{f\mathrm{v} n}/N_{f\mathrm{v} 0}$, in addition to the $x$ scaling 
mentioned above.

Also, when a sea quark is knocked out of a hadron, it must leave behind a
corresponding antisea parton in the beam remnant. We call this a companion 
quark. In the perturbative approximation the sea quark 
$\mathrm{q}_{\mathrm{s}}$ and its companion $\mathrm{q}_{\mathrm{c}}$ 
come from a gluon branching
$\mathrm{g} \to \mathrm{q}_{\mathrm{s}} + \mathrm{q}_{\mathrm{c}}$ 
(it is implicit that if $\mathrm{q}_{\mathrm{s}}$ is a quark, 
$\mathrm{q}_{\mathrm{c}}$ is its antiquark). Starting from this 
perturbative ansatz, and neglecting other interactions and 
any subsequent perturbative evolution of the $\mathrm{q}_{\mathrm{c}}$, 
we obtain the $q_{\mathrm{c}}$ distribution from
the probability that a sea quark $\mathrm{q}_{\mathrm{s}}$, carrying a
momentum fraction $x_{\mathrm{s}}$, is produced by the branching of a
gluon with momentum fraction $y$, so that the 
companion has a momentum fraction $x=y-x_{\mathrm{s}}$, 
\begin{equation}
q_{\mathrm{c}}(x;x_{\mathrm{s}}) \propto \int_0^1 g(y) \, 
P_{\mathrm{g}\to\mathrm{q}_{\mathrm{s}}\mathrm{q}_{\mathrm{c}}}(z) \, 
\delta(x_{\mathrm{s}}-zy)~\mathrm{d} z =
\frac{g(x_{\mathrm{s}}+x)}{x_{\mathrm{s}}+x} \, 
P_{\mathrm{g}\to\mathrm{q}_{\mathrm{s}}\mathrm{q}_{\mathrm{c}}}
\left(\frac{x_{\mathrm{s}}}{x_{\mathrm{s}}+x}\right), 
\end{equation}
with $P_{\mathrm{g}\to\mathrm{q}_{\mathrm{s}}\mathrm{q}_{\mathrm{c}}}$ 
the usual DGLAP gluon splitting kernel. A simple ansatz 
$g(x) \propto (1-x)^n/x$ is here used for the gluon. Normalizations
are fixed so that a sea quark has exactly one companion.
Qualitatively, $xq_{\mathrm{c}}(x;x_s)$ is peaked around 
$x \approx x_{\mathrm{s}}$, by virtue of the symmetric 
$P_{\mathrm{g}\to\mathrm{q}_{\mathrm{s}}\mathrm{q}_{\mathrm{c}}}$ 
splitting kernel.

Without any further change, the reduction of the valence
distributions and the introduction of companion distributions, in the 
manner described above, would result in a violation of the total 
momentum sum rule, that the $x$-weighted parton densities should 
integrate to $X$: by removing a valence quark from the parton 
distributions we also remove a total amount of momentum corresponding 
to $\langle x_{f\mathrm{v}} \rangle$, the average momentum fraction 
carried by a valence quark of flavour $f$,
\begin{equation}
\langle x_{f\mathrm{v} n} \rangle \equiv 
\frac{\int_0^X xq_{f\mathrm{v} n}(x,Q^2)~\mathrm{d} x}%
{\int_0^X q_{f\mathrm{v} n}(x,Q^2)~\mathrm{d} x} = X \, 
\langle x_{f\mathrm{v} 0} \rangle ~,
\end{equation}
and by adding a companion distribution we add an analogously defined
momentum fraction. 

To ensure that the momentum sum rule is still respected, we assume that 
the sea+gluon normalizations fluctuate up when a valence distribution is 
reduced and down when a companion distribution is added, by a 
multiplicative factor
\begin{equation}
a = \frac{1-\sum_fN_{f\mathrm{v} n}\langle x_{f\mathrm{v} 0} \rangle
-\sum_{f,j} \langle x_{f\mathrm{c}_j 0} \rangle}{1- \sum_fN_{f\mathrm{v} 0}
\langle x_{f\mathrm{v} 0} \rangle} ~. 
\end{equation}
The requirement of a physical $x$ range is of course still maintained by 
`squeezing' all distributions into the interval $x\in[0,X]$. The full parton 
distributions after $n$ interactions thus take the forms
\begin{eqnarray}
q_{f n}\left(x, Q^2\right) & = &  \frac{1}{X} \left[ 
\frac{N_{f\mathrm{v} n}}{N_{f\mathrm{v} 0}}
{q_{f\mathrm{v} 0}\left(\frac{x}{X}, Q^2\right)} + 
a\, q_{f \mathrm{s} 0} \left(\frac{x}{X}, Q^2\right) + 
\sum_{j} q_{f\mathrm{c}_j} \left(\frac{x}{X};x_{s_j}\right) \right]  ~,
\\
\displaystyle {g_n(x)} &=&\frac{a}{X} g_0\left(\frac{x}{X}, Q^2\right) ~,
\end{eqnarray}
where $q_{f\mathrm{v} 0}$ ($q_{f \mathrm{s}0}$) denotes the original 
valence (sea) distribution of flavour $f$, and the index $j$ on the 
companion distributions $q_{f\mathrm{c}_j}$ counts different companion 
quarks of the same flavour $f$. 

After the perturbative interactions have taken each their fraction of
longitudinal momentum, the remaining momentum is to be shared between 
the beam remnant partons. Here, valence quarks receive an $x$ picked at 
random according to a small-$Q^2$ valence-like parton density, while sea 
quarks must be companions of one of the initiator quarks, and hence should 
have an $x$ picked according to the $q_{\mathrm{c}}(x ; x_{\mathrm{s}})$ 
distribution introduced above. In the rare case that no valence quarks 
remain and no sea quarks need be added for flavour conservation, the beam 
remnant is represented by a gluon, carrying all of the beam remnant 
longitudinal momentum. 

Further aspects of the model include the possible formation of composite
objects in the beam remnants (e.g.\ diquarks) and the addition
of non-zero primordial $k_{\perp}$ values to the parton shower
initiators. Especially the latter introduces some complications, to
obtain consistent kinematics. Details on these aspects 
are presented in \cite{multintinprep}. 

\section{COLOUR CORRELATIONS}

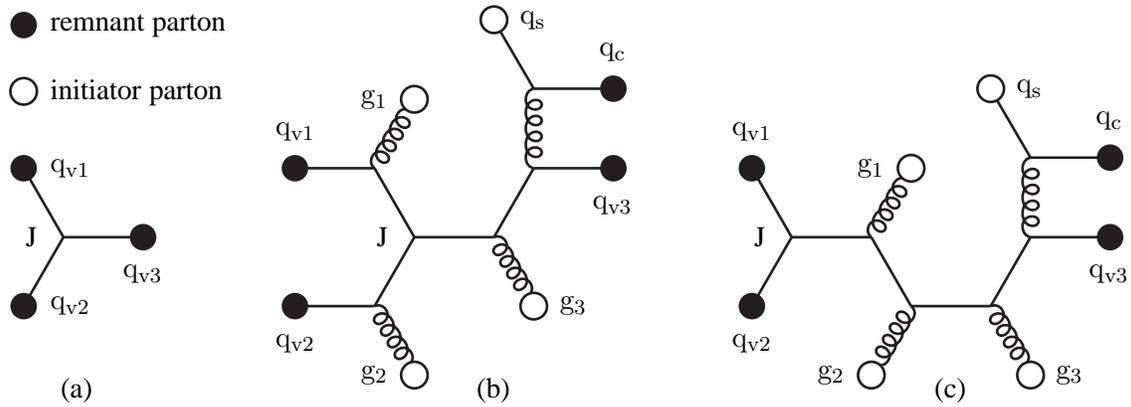
\begin{figure}[t]
\begin{center}
\begin{picture}(120,150)(-30,-60)
\SetWidth{1}
\Text(-10,0)[r]{J}
\Vertex(-15,80){5}\Text(-5,80)[l]{remnant parton}
\BCirc(-15,55){5}\Text(-5,55)[l]{initiator parton}
\Line(0,0)(30,0)\Vertex(30,0){5}
\Text(30,-14)[]{$\mathrm{q}_{\mathrm{v}3}$}
\Line(0,0)(-15,26)\Vertex(-15,26){5}
\Text(-5,26)[l]{$\mathrm{q}_{\mathrm{v}1}$}
\Line(0,0)(-15,-26)\Vertex(-15,-26){5}
\Text(-5,-26)[l]{$\mathrm{q}_{\mathrm{v}2}$}
\Text(5,-58)[]{(a)}
\end{picture}
\begin{picture}(150,150)(-40,-60)
\SetWidth{1}
\Text(-10,0)[r]{J}
\Line(0,0)(-15,26)
\Line(-15,26)(-45,26)\Vertex(-45,26){5}
\Text(-45,40)[]{$\mathrm{q}_{\mathrm{v}1}$}
\Gluon(-15,26)(-2,48){3}{4}\BCirc(0,52){5}
\Text(-10,52)[r]{$\mathrm{g}_1$}
\Line(0,0)(-15,-26)
\Line(-15,-26)(-45,-26)\Vertex(-45,-26){5}
\Text(-45,-40)[]{$\mathrm{q}_{\mathrm{v}2}$}
\Gluon(-15,-26)(-2,-48){3}{4}\BCirc(0,-52){5}
\Text(-10,-52)[r]{$\mathrm{g}_2$}
\Line(0,0)(30,0)
\Gluon(30,0)(43,-22){3}{4}\BCirc(45,-26){5}
\Text(55,-26)[l]{$\mathrm{g}_3$}
\Line(30,0)(45,26)
\Line(45,26)(75,26)\Vertex(75,26){5}
\Text(75,12)[]{$\mathrm{q}_{\mathrm{v}3}$}
\Gluon(45,26)(45,56){3}{4}
\Line(45,56)(75,56)\Vertex(75,56){5}
\Text(75,70)[]{$\mathrm{q}_{\mathrm{c}}$}
\Line(45,56)(31,80)\BCirc(30,82){5}
\Text(40,82)[l]{$\mathrm{q}_{\mathrm{s}}$}
\Text(30,-58)[]{(b)}
\end{picture}
\begin{picture}(160,150)(-30,-60)
\SetWidth{1}
\Text(-10,0)[r]{J}
\Line(0,0)(-15,26)\Vertex(-15,26){5}
\Text(-15,40)[]{$\mathrm{q}_{\mathrm{v}1}$}
\Line(0,0)(-15,-26)\Vertex(-15,-26){5}
\Text(-15,-40)[]{$\mathrm{q}_{\mathrm{v}2}$}
\Line(0,0)(30,0)
\Gluon(30,0)(43,22){3}{4}\BCirc(45,26){5}
\Text(35,26)[r]{$\mathrm{g}_1$}
\Line(30,0)(45,-26)
\Gluon(45,-26)(32,-48){3}{4}\BCirc(30,-52){5}
\Text(20,-52)[r]{$\mathrm{g}_2$}
\Line(45,-26)(75,-26)
\Gluon(75,-26)(88,-48){3}{4}\BCirc(90,-52){5}
\Text(100,-52)[l]{$\mathrm{g}_3$}
\Line(75,-26)(90,0)
\Line(90,0)(120,0)\Vertex(120,0){5}
\Text(120,-14)[]{$\mathrm{q}_{\mathrm{v}3}$}
\Gluon(90,0)(90,30){3}{4}
\Line(90,30)(120,30)\Vertex(120,30){5}
\Text(120,44)[]{$\mathrm{q}_{\mathrm{c}}$}
\Line(90,30)(76,54)\BCirc(75,56){5}
\Text(85,56)[l]{$\mathrm{q}_{\mathrm{s}}$}
\Text(60,-58)[]{(c)}
\end{picture}
\end{center}
\caption{(a) The initial state of a baryon, with the valence quarks
colour-connected via a central string junction J. 
(b) Example of a topology with initiators connected at random.
(c) Alternative with the junction in the remnant.}
\label{fig:initialstate}
\end{figure}

The initial state of a baryon may be represented by three valence quarks,
connected antisymmetrically in colour via a central junction, which acts 
as a switchyard for the colour flow and carries the net baryon number, 
Fig.~\ref{fig:initialstate}a. 

The colour-space evolution of this state into the initiator and remnant 
partons actually found in a given event is not predicted by perturbation 
theory, but is crucial in determining how the system hadronizes; in the 
Lund string model \cite{Andersson:1983ia}, two colour-connected final 
state partons together define a string piece, which hadronizes by 
successive non-perturbative breakups along the string. Thus, the colour 
flow of an event determines the topology of the hadronizing strings, 
and consequently where and how~many hadrons will be produced. 
The question can essentially be reduced to one of choosing a fictitious
sequence of gluon emissions off the initial valence topology, since sea 
quarks together with their companion partners are associated with parent 
gluons, by construction.

The simplest solution is to assume that gluons are attached to the initial
quark lines in a random order, see Fig.~\ref{fig:initialstate}b. 
If so, the junction would rarely be colour-connected directly to two 
valence quarks in the beam remnant, and the initial-state baryon number 
would be able to migrate to large $p_{\perp}$ and small $x_F$ values.
While such a mechanism should be present, there are reasons to believe
that a purely random attachment exaggerates the migration effects.  
Hence a free parameter is introduced to suppress gluon attachments onto 
colour lines that lie entirely within the remnant, so that topologies
such as Fig.~\ref{fig:initialstate}c become more likely. 

This still does not determine the order in which gluons are attached to 
the colour line between a valence quark and the junction. We consider a 
few different possibilities: 1) random, 2) gluons are ordered according 
to the rapidity of the hard scattering subsystem they are associated with, 
and 3) gluons are ordered so as to give rise to the smallest possible 
total string lengths in the final state. The two latter possibilities 
correspond to a tendency of nature to minimize the total potential 
energy of the system, i.e.\ the string length. Empirically such a 
tendency among the strings formed by multiple interactions is supported 
e.g.\ by the observed rapid increase of 
$\langle p_{\perp} \rangle$ with $n_{\mathrm{charged}}$. It appears, 
however, that a string minimization in the initial state is not enough,
and that also the colours inside the initial-state cascades and hard 
interactions may be nontrivially correlated. These studies are still 
ongoing, and represent the major open issues in the new model. 

\section{CONCLUSION}

A new model for the underlying event in hadron--hadron collisions 
\cite{multintinprep} has been introduced. This model extends the multiple 
interactions mechanism proposed in \cite{Sjostrand:1987su} with the
possibility of non-trivial flavour and momentum correlations, with
initial- and final-state showers for all interactions, and with several 
options for colour correlations between initiator and remnant partons. 
Many of these improvements rely on the development of junction 
fragmentation in \cite{Sjostrand:2002ip}. 

This is not the end of the line. Rather we see that many issues remain
to understand better, such as colour correlations between partons
in interactions and beam remnants, whereas others have not yet been 
studied seriously, such as the extent to which two interacting partons
stem from the same initiator. Theoretical advances alone cannot solve
all problems; guidance will have to come from experimental information. 
The increased interest in such studies bodes well for the future.

\bibliography{biblifile}

\end{document}